\documentclass[12pt]{article}
\usepackage[utf8]{inputenc}
\usepackage{amsmath,graphicx}
\usepackage[style=vancouver]{biblatex}
\usepackage{fullpage}
\usepackage{lineno}
\usepackage{xcolor}
\usepackage{pdflscape}
\usepackage{hyperref}
\usepackage{authblk}
\usepackage{amssymb}

\usepackage{dcolumn}
\usepackage{bm}
\usepackage{textcomp}

\title{\sc{Likelihood-based inference, identifiability and prediction using count data from lattice-based random walk models}}

\author[1]{Yihan Liu}
\author[1]{David J. Warne}
\author[1]{Matthew J. Simpson}
\affil[1]{School of Mathematical Sciences, Queensland University of Technology, Brisbane, Australia.}
\date{}

\begin{document}


\maketitle

\begin{abstract}
\textit{In vitro} cell biology experiments are routinely used to characterize cell migration properties under various experimental conditions.  These experiments can be interpreted using lattice-based random walk models to provide insight into underlying biological mechanisms, and continuum limit partial differential equation (PDE) descriptions of the stochastic models can be used to efficiently explore model properties instead of relying on repeated stochastic simulations.  Working with efficient PDE models is of high interest for parameter estimation algorithms that typically require a large number of forward model simulations.  Quantitative data from cell biology experiments usually involves non-negative cell counts in different regions of the experimental images, and it is not obvious how to relate finite, noisy count data to the solutions of continuous PDE models that correspond to noise-free density profiles.  In this work we illustrate how to develop and implement likelihood-based methods for parameter estimation, parameter identifiability and model prediction for lattice-based models describing collective migration with an arbitrary number of interacting subpopulations.   We implement a standard additive Gaussian measurement error model as well as a new physically-motivated multinomial measurement error model that relates noisy count data with the solution of continuous PDE models. Both measurement error models lead to similar outcomes for parameter estimation and parameter identifiability, whereas the standard additive Gaussian measurement error model leads to non-physical prediction outcomes.  In contrast, the new multinomial measurement error model involves a lower computational overhead for parameter estimation and identifiability analysis, as well as leading to physically meaningful model predictions. Open access Julia software required to replicate the results in this study are available on \href{https://github.com/YiH-Liu/Likelihood-based-inference.git}{GitHub}.
\end{abstract}

\section{Introduction} \label{sec:Intro}
\begin{figure}[htp!]
\centering
\includegraphics[width=1\textwidth]{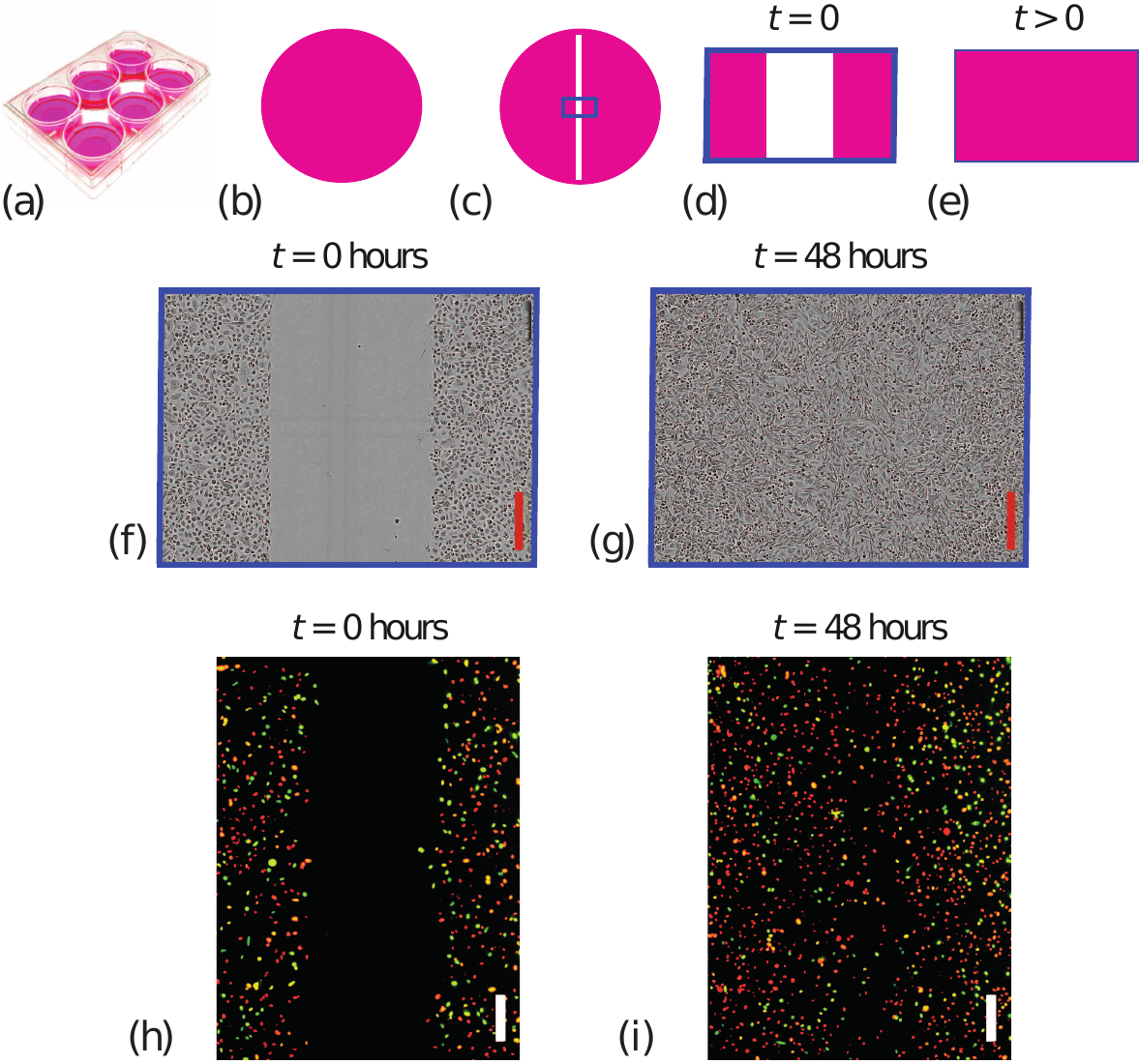}
\caption{(a)–(e) Scratch assay schematic. (a)–(b) Cell monolayers are grown in 6-well tissue culture plates, where each well within the tissue culture plate has a diameter of 35 mm. (c) The scratched monolayer is superimposed with the imaging field of view (solid blue). (d)–(e) Schematic showing population invasion over time within the field of view (solid blue). (f)-(g) Experimental images from a scratch assay using a single population of prostate cancer cells at time $t=0$ and $t=48$ hours. The red scale bar corresponds to 300 $\text{\textmu m}$~\cite{jin2016reproducibility}. (h)-(i) Experimental images from a scratch assay for using a more complicated population of melanoma cells composed of two populations indicated by red and green fluorescence at time $t=0$ and $t=48$ hours. The red scale bar corresponds to 200 $\text{\textmu m}$~\cite{vittadello2018mathematical}. All images are reused with permission.}
\label{fig:1}
\end{figure}

Two-dimensional cell migration assays, also called scratch assays, are routinely used to quantify how different populations of cells migrate~\cite{liang2007vitro}.  These simple experiments, outlined in Figure \ref{fig:1}, can be used to study potential drug treatments~\cite{grada2017research}, different nutrient availability conditions~\cite{Khain2011}, or interactions between different populations of cells~\cite{Falco2023}.  Scratch assays involve growing uniform monolayers of cells on tissue culture plates before part of the monolayer is \textit{scratched} away using a sharp-tipped instrument like a razor blade.  The resulting recolonisation of the scratched region is then imaged, and the rate at which the scratch closes provides a simple measure of the ability of cells to migrate~\cite{liang2007vitro,grada2017research}.  Scratch assays conducted over relatively short periods of time (e.g. less than 24 hours) are often used to focus on cell migration~\cite{johnston2014}, whereas experiments conducted over longer periods of time (e.g. greater than 24 hours) provide insight into the role of cell migration and cell proliferation combined~\cite{johnston2014}.

One way of interpreting scratch assays is to implement a stochastic random walk model describing the motion of individual agents on a lattice~\cite{Khain2011,simpson2010cell,Callaghan2006,ross2017using,Browning2017}.  The location of agents can be initialised to mimic the initial geometry of the scratch, and then agents are allowed to undergo a random walk, either biased or unbiased, with a simple exclusion mechanism to model crowding effects.  Carefully choosing parameters in the random walk model to replicate experimental observations provides biological insight into the roles of directed and undirected migration among different subpopulations of cells within the experiment.   Using a stochastic simulation model to interpret these experiments is advantageous because stochastic models allow us to keep track of individual cells within the population, as well as capturing the role of stochasticity in the experiments~\cite{vo2015quantifying}.

In the last decade there has been an increasing interest in parameter estimation for stochastic models of cell migration experiments~\cite{ross2017using,Browning2017,vo2015quantifying}.  When working with a discrete model of a spatially explicit biological process, such as cell migration, it has become customary to implement some form of Approximate Bayesian Computation (e.g. ABC rejection, ABC-MCMC)~\cite{ross2017using,Browning2017,vo2015quantifying,Vo2015b},  which is often motivated by noting that stochastic models are not associated with a tractable likelihood function.  A similar trend has emerged in the spatial ecology literature, where parameter estimation has been dominated by different ABC methods~\cite{bode2020,Hamilton2005,Rasmussen2012,Vaart2015,Vaart2018}.  In this work we demonstrate how to take a different approach by using likelihood-based methods for parameter estimation, parameter identifiability and model prediction~\cite{wasserman2004all}. Noting that scratch assays are often quantified by reporting cell counts in different spatial regions of experimental images~\cite{jin2016reproducibility,vittadello2018mathematical,gnerucci2020scratch}, we explore how to use approximate continuum limit PDE models as a computationally efficient process model to describe the key mechanistic features in a scratch assay~\cite{simpson2024}.  We work in a general setting by considering migration and crowding of a population composed of $s=1,2,3,\ldots,S$ potentially distinct subpopulations of agents.  

The main focus of our work is to compare two approaches for relating the solution of the PDE model to the observed noisy count data: (i) a traditional additive Gaussian measurement error model; and (ii) a new physically-motivated multinomial measurement error model.  It is worth noting that applications of parameter estimation in biological applications involve working with an additive Gaussian measurement error model, and that this choice is often implemented without critically examining the implications of this assumption~\cite{hines2014determination,simpson2020}.  Using maximum likelihood estimation and the profile likelihood~\cite{pawitan2001,raue2009,raue2013,villaverde2019,villaverde2022}, we obtain best-fit parameter estimates and likelihood-based confidence intervals that allow us to examine the practical identifiability of parameter estimates derived from noisy count data.  Importantly, we also use a likelihood-based method to map the variability in parameter confidence sets to explore the variability in model predictions~\cite{murphy2024implementing}.  While both the additive Gaussian and multinomial measurement error models perform similarly in terms of parameter estimation and parameter identifiability, we show that the standard additive Gaussian measurement error model leads to unphysical predictions of negative agent counts, or counts of agents that locally exceed the maximum carrying capacity of the lattice.  In contrast, the new physically-motivated multinomial measurement error model is simpler and faster to implement computationally compared to usual additive Gaussian measurement error model.  Furthermore, the multinomial measurement error model leads to physically realistic model predictions while also avoiding the computational expense of a more standard ABC approach using a far more expensive stochastic process model.

This manuscript is organised as follows.  In Section \ref{sec:Methods} we introduce the stochastic simulation model and the associated continuum-limit PDE.  We also provide a general description of how noisy count data can be generated using the stochastic simulation algorithm to mimic experimental measurements for problems involving both single populations and two subpopulations of distinct agents.  The two measurement error models are introduced along with methods for maximum likelihood estimation, practical identifiability analysis using the profile likelihood and likelihood-based prediction.  In section \ref{sec:Results} we present specific results for parameter estimation, parameter identifiability analysis and model prediction for a simple case of count data with one population of agents and a more realistic case of count data associated with a population composed of two subpopulations of agents.  Finally, in Section \ref{sec:Conclusion} we summarise our findings and outline opportunities for future work.

\section{Methods}\label{sec:Methods}
\subsection{Mathematical Models}\label{sec:Mathematical model}
In this study, we use two types of mathematical modelling frameworks. First, we use a stochastic lattice-based random walk model that will be described in Section~\ref{sec:sm}. The motivation for using a computationally expensive stochastic model is that it provides a high-fidelity means of generating noisy data that is compatible with the kind of data generated experimentally. Second, we use a computationally efficient continuum limit description of the stochastic model.  The continuum limit description takes the form of a system PDEs that will be described in Section~\ref{sec:cm}.

\subsubsection{Stochastic model} \label{sec:sm}
Experimental images from scratch assays shown in Figure~\ref{fig:1} motivate our stochastic model. Scratch assay experiments are routinely used in experimental cell biology to quantify cell migration. For example,  these experiments are often used to explore how different surface coatings or different putative drugs impact the ability of cells to migrate.  Scratch assays are performed by growing a uniform population of cells in a tissue culture plate.  A sharp-tipped instrument is used to create a scratch within the uniform monolayer, and individual cells within the population undergo random migration, where the motility of individual cells is influenced by crowding effects.  The net outcome of this random migration is that cells move into the scratched region, and this closes the scratched region over time.  

We use a lattice-based stochastic model to simulate the migration of $S$ distinct populations of agents on a two-dimensional lattice with lattice spacing $\Delta$~\cite{Simpson2009}.   In our simulations we think of each agent representing an individual cell in the experiment.  The size of the lattice is $W \times H$, where $W$ is the width of the lattice and $H$ is the height of the lattice. Each site is indexed $(i,j)$, and each site is associated with a unique Cartesian coordinate $(x_i, y_j)$, where $x_i = (i - 1) \Delta$ with $i = 1, 2, 3, \ldots, I$ and $y_j = (j - 1) \Delta$ with $j = 1, 2, 3, \ldots, J$.

Stochastic simulations are performed using a random sequential update method~\cite{Chowdhury2005}. Simulations are initialised by placing a total of $N$ agents from $S$ distinct agent subpopulations on the lattice. If $N_{s}$ is the number of agents in the $s$th subpopulation, then we have $N = \displaystyle{\sum_{s=1}^{S} N_{s}}$.  To evolve the stochastic algorithm from time $t$ to time $t+\tau$, $N$ agents are selected at random, one at a time, with replacement and given an opportunity to move~\cite{Simpson2009}.  When an agent belonging to subpopulation $s$ is chosen, that agent attempts to move with probability $P_s \in [0,1]$.  The target site for potential motility events is chosen in the following way. The probability that a motile agent at site $(i,j)$ attempts to move to site $(i,j\pm1)$ is $1/4$, whereas the probability that a motile agent at site $(i,j)$ attempts to move to site $(i \pm 1,j)$ is $(1 \pm \rho_s)/4$.  Here, $ \rho_{s} \in [-1,1] $ is a bias parameter that influences the left-right directional bias in the horizontal direction.  Setting  $ \rho_{s} = 0 $ indicates that there is no bias for agents in the $s$th subpopulation.  Any potential motility event that would place an agent on an occupied site is aborted. This means that our stochastic simulation algorithm is closely related to an exclusion process~\cite{Chowdhury2005}.  Periodic boundary conditions are applied along the horizontal boundaries, and reflecting boundary conditions are imposed along the vertical boundaries.  Time steps in the simulations are indexed by $k$ so that $t = k\tau$ for $k=1,2,3,\ldots$. To keep our simulation framework general we always work with a dimensionless simulations by setting $\Delta = \tau = 1$.  These simulations can be re-scaled using appropriate length and time scales to accommodate cells of different sizes and motility rates~\cite{simpson2010cell}. 

Although cells in the experiments in Figure~\ref{fig:1} are free to move in any direction, scratching a uniform monolayer along the vertical direction means that the density of cells is independent of the vertical location within the image, and that the macroscopic density of cells varies with horizontal position.  Therefore, these experimental images have been quantified by superimposing a series of uniformly-spaced columns across each image and counting the number of cells within each column~\cite{jin2016reproducibility,vittadello2018mathematical,gnerucci2020scratch}.  This approach summarises the outcome of each experiment as a series of count data as a function of horizontal location of each column.  The most standard way of reporting outcomes of a scratch assay is to image the experiment once at the end of the experiment and then counts of cells can be determined and reported.  Alternatively, it is possible to repeat the imaging and counting process across a number of time points and report a time series of count data.    Our modelling framework can be applied to either approach, but we will present our results for the more standard approach of working with data at one time point only, and in Section \ref{sec:Conclusion} we will explain how our methodology generalises to working with a time series of count data. 

We will now describe how the stochastic random walk model can be used to generate count data in exactly the same way as count data are generated experimentally.  In this work we always consider stochastic simulations that mimic the same geometry and design as in the experimental scratch assays. Therefore all simulations are initialised so that the expected occupancy status of lattice sites within the same column of the lattice is identical. Together with the boundary conditions, this ensures that the occupancy status of any lattice site is, on average, independent of vertical location at any time during the simulation.

The outcome of the stochastic simulations is to determine the occupancy of each lattice site for different  subpopulations as a function of time. To quantify this we let  $C_{s}^{\star}(i,j,k)$ denote the occupancy of site $(i,j)$, for subpopulation $s$ after $k$ time steps. If site $(i,j)$ is occupied by an agent from subpopulation $s$ we have $C_{s}^{\star}(i,j,k)=1$, otherwise $C_{s}^{\star}(i,j,k)=0$. With this notation, the observed total agent count from subpopulation $s$ in column $i$ after $k$ time steps is 
\begin{equation}
\label{eq:data}
C_{s}^{\text{o}}(i,k) = \sum_{j=1}^{J} C_{s}^{\star}(i,j,k).
\end{equation}
These counts are bounded since $C_{s}^{\text{o}}(i,k) \in [0,J]$ for all $s=1,2,3,\ldots, S$.  

\begin{figure}[htp!]
\centering
\includegraphics[width=1\textwidth]{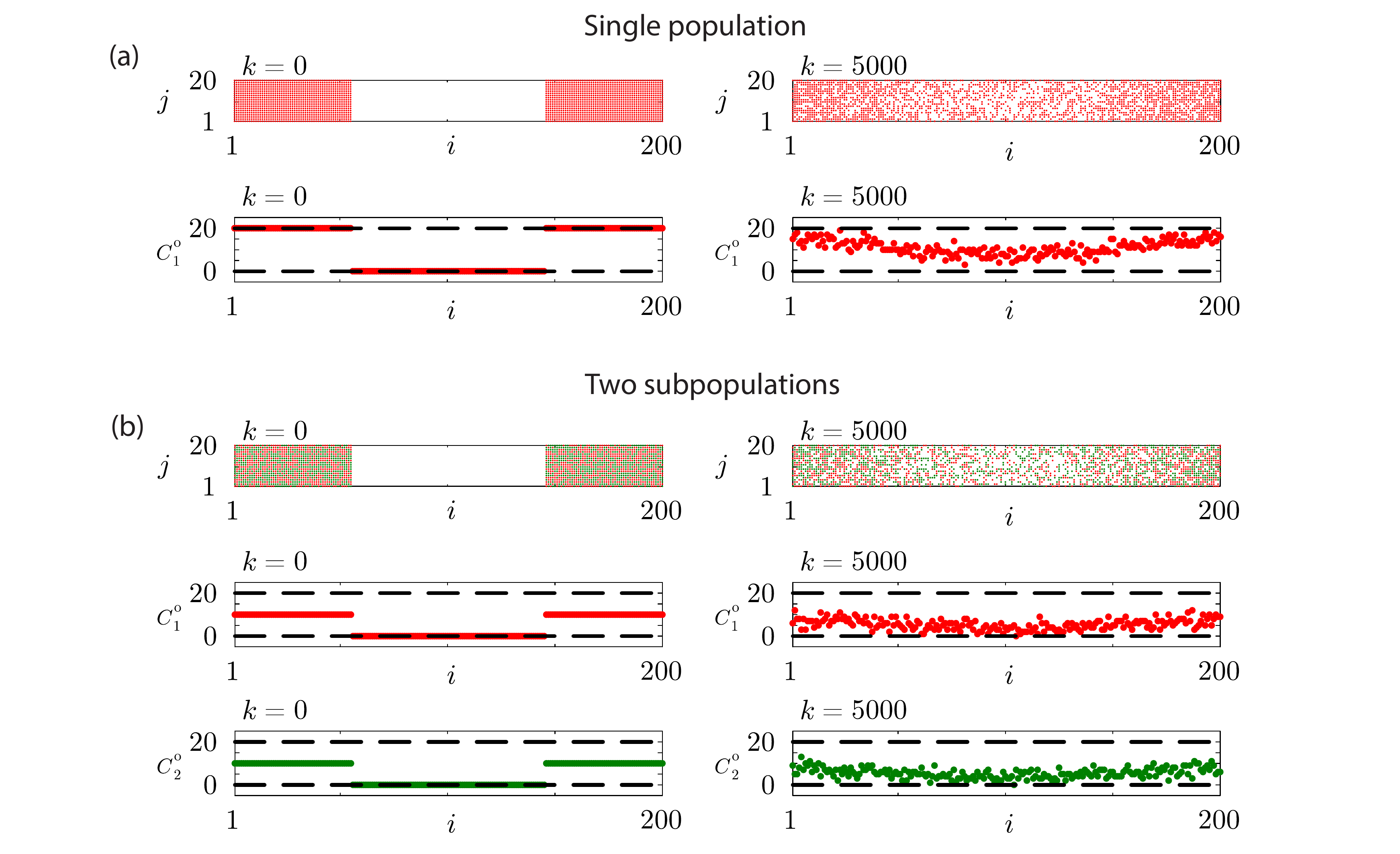}
\caption{Motivating stochastic simulations on a $200 \times 20$ lattice illustrate agent snapshots and count data for simulations involving a single population and populations composed of two interacting subpopulations.  Results in (a) correspond to a single population with $(P_{1},\rho_1)^\top = (1,0)^\top$.   The top row shows simulation snapshots at $k=0$ and $k=5000$, with the lower row showing the corresponding count data, $C_{1}^{\text{o}}(i,k)$  for $i=1,2,3,\dots,200$. Results in (b) correspond to a population composed of two interacting subpopulations with $(P_1,P_2,\rho_1,\rho_2)^\top = (1,1,0,0)^\top$.   The top row shows the simulation snapshots at $k=0$ and $k=5000$ where agents from subpopulation 1 are red and agents from subpopulation 2 are green.  The middle row shows the corresponding count data for subpopulation 1, $C_{1}^{\text{o}}(i,k)$ for $i=1,2,3,\dots,200$. The lower row shows the corresponding count data for subpopulation 2, $C_{2}^{\text{o}}(i,k)$ for $i=1,2,3,\dots,200$.}
\label{fig:2}
\end{figure}

To mimic the scratch assay experiment in Figure~\ref{fig:1}(f)-(g) with a single population we consider stochastic simulations with $ S=1 $ on a $ 200 \times 20 $ lattice.  Initially the lattice is populated so that each site with $ 1 \leq i \leq 55 $ and $ 146 \leq i \leq 200 $ is fully occupied.  Agents move without directional bias by setting $P_1=1$ and $\rho_1=0$.  Results in Figure~\ref{fig:2}(a) show snapshots of the distribution of agents together with plots of the associated count data at the beginning and conclusion of the simulation.  It is relevant to note that the count data at the end of the simulation, $ C_{1}^{\text{o}}(i,k) $ for $i=1,2,3,\ldots,I$, is noisy and exhibits large fluctuations across the columns of the lattice.  These fluctuations are similar to the fluctuations observed in count data generated by quantifying images from a scratch assay performed using a single population of cells such as the experimental images in Figure~\ref{fig:1}(f)-(g).

We now explore how to use the stochastic model to mimic the scratch assay experiment in Figure~\ref{fig:1}(h)-(i) with two subpopulations of cells  by initialising a $200 \times 20$ lattice with two subpopulations of agents, $S=2$. Stochastic simulations are initialised by randomly populating 50$\%$ of sites in each column with $1 \leq i \leq 55$ and $146\leq i \leq 200$ with agents from subpopulation 1.  The remaining sites in these columns are occupied by agents from subpopulation 2.  Simulations are performed with $P_1=P_2=1$ and $\rho_1=\rho_2=0$.  For this choice of parameters it turns out that both subpopulations have the same migration rate and bias parameter which means that both subpopulations behave identically.  Later in Section \ref{sec:data} we will show that the same ideas apply when the subpopulations are characterised by different migration rates and bias parameters.  Results in Figure~\ref{fig:2}(b) show snapshots of the distribution of agents and plots of the associated count data.  Since we are working with two subpopulations of agents we are able to generate two sets of count data per column, $C_{1}^{\text{o}}(i,k)$ and $C_{2}^{\text{o}}(i,k)$ for $i=1,2,3,\ldots,I$, after $k$ time steps.  In this case we see that the count data at the end of the simulation are noisy and exhibit large fluctuations. 

\subsubsection{Continuum model} \label{sec:cm}
The stochastic model presented in Section \ref{sec:sm} is well-suited to mimic noisy data from biological experiments. However, the stochastic model is computationally expensive, which means that it is not well-suited for parameter estimation, which can require a very large number of forward simulations. To make progress we will use a computationally efficient continuum limit approximations of the stochastic model. The continuum limit description of the stochastic model is given by a system of PDEs~\cite{Simpson2009} that can be written in conservation form as 
\begin{equation}
\label{eq:cont}
\frac{\partial c_{s}}{\partial t} = -\frac{\partial \mathcal{J}_{s}}{\partial x}, \quad \textrm{for} \quad  s = 1,2,3, \dots, S,
\end{equation}
where $c_{s}(x, t)$ is the dimensionless density of population $s$ at location $x$ and time $t$, and the flux of subpopulation $s$ can be written as
\begin{equation}
\label{eq:Js}
\mathcal{J}_{s} = -D_{s} \left(1-\sum_{s=1}^{S} c_{s}\right)\frac{\partial c_{s}}{\partial x} - D_{s}c_{s}\frac{\partial}{\partial x}\left(\sum_{s=1}^{S} c_{s}\right) + v_{s}c_{s} \left(1-\sum_{s=1}^{S} c_{s}\right),
\end{equation}
where
\begin{equation}
\label{eq:Dv}
D_{s} = \lim_{\Delta, \tau \to 0} \frac{P_{s}\Delta^2}{4\tau} \quad \quad \text{and} \quad \quad v_{s} = \lim_{\Delta, \tau \to 0} \frac{P_{s}\rho_{s}\Delta}{2\tau}.
\end{equation}

In this study we focus on applications involving either one or two subpopulations. For the single population model $S=1$, the continuum model simplifies to
\begin{equation}
\label{eq:cont_1p}
\begin{aligned}
\frac{\partial c_{1}}{\partial t}&= - \frac{\partial  \mathcal{J}_{1}}{\partial x}, \quad \text{with} \quad \mathcal{J}_{1}&= -D_{1}\frac{\partial c_{1}}{\partial x} + v_{1}c_{1}\left(1-c_{1}\right),
\end{aligned}
\end{equation}
where
\begin{equation}
\label{eq:Dv_1p}
D_{1} = \lim_{\Delta, \tau \to 0} \frac{P_{1}\Delta^2}{4\tau}  \quad \quad \text{and} \quad \quad v_{1} = \lim_{\Delta, \tau \to 0} \frac{P_{1}\rho_{1}\Delta}{2\tau}.
\end{equation}
For two subpopulations, $S = 2$, the continuum model simplifies to
\begin{equation}
\label{eq:cont_2p}
\begin{aligned}
\frac{\partial c_{1}}{\partial t}&= - \frac{\partial  \mathcal{J}_{1}}{\partial x}, \quad \quad 
 \text{and} \quad \quad \frac{\partial c_{2}}{\partial t}&= -\frac{\partial  \mathcal{J}_{2}}{\partial x}, 
\end{aligned}
\end{equation}
with
\begin{equation}
\label{eq:J1}
\begin{aligned}
 \mathcal{J}_{1}&= -D_{1}\left(1-c_{1}-c_{2}\right)\frac{\partial c_{1}}{\partial x}-D_{1}c_{1}\frac{\partial}{\partial x}\left(c_{1}+c_{2}\right)+v_{1}c_{1}\left(1-c_{1}-c_{2}\right),
\end{aligned}
\end{equation}
\begin{equation}
\label{eq:J2}
\begin{aligned}
 \mathcal{J}_{2}&= -D_{2}\left(1-c_{1}-c_{2}\right)\frac{\partial c_{2}}{\partial x}-D_{2}c_{2}\frac{\partial}{\partial x}\left(c_{1}+c_{2}\right)+v_{2}c_{2}\left(1-c_{1}-c_{2}\right),
\end{aligned}
\end{equation}
where
\begin{equation}
\label{eq:Dv_2p}
\begin{aligned}
D_{1} = \lim_{\Delta, \tau \to 0} \dfrac{P_1 \Delta^2}{4\tau}, \quad v_1 = \lim_{\Delta, \tau \to 0} \dfrac{P_1 \rho_1 \Delta}{2\tau}, \quad D_{2} = \lim_{\Delta, \tau \to 0} \frac{P_2 \Delta^2}{4\tau}\quad 
\text{and} \quad v_2 = \lim_{\Delta, \tau \to 0} \dfrac{P_2 \rho_2 \Delta}{2\tau}.
\end{aligned}
\end{equation}

An implicit assumption in the derivation of the continuum limit model is that we are dealing with a sufficiently large lattice such that fluctuations in count data are negligible.  In this idealised scenario it is possible to relate count data to the solution of the continuum limit model by
\begin{align}\label{eq:CtmDiscrete}
c_{s}(x_i,t)  &= \lim_{J \to \infty} \left[\dfrac{\sum_{j=1}^{J}C_{s}^{\star}(i,j,k)}{J} \right], \notag \\
&= \lim_{J \to \infty} \left[\dfrac{C_{s}^{\text{o}}(i,k)}{J}\right],
\end{align}
where $x_i$ is the central position of the $i$th column and $C_{s}^{\star}(i,j,k)$ indicates the occupancy of lattice site $(i,j)$ for subpopulation $s$ after $k$ time steps of the stochastic model.   This relationship, which has been verified computationally~\cite{Simpson2009}, indicates that the solution of the continuum limit PDE model approaches the column-averaged density estimates obtained using count data only in the impractical situation where the height of the lattice (or the height of the experimental image) is sufficiently large.  In practice, however, experiments always involve relatively small fields of view and consequently the associated count data can involve relatively large fluctuations that are similar to our simulation-derived count data in Figure \ref{fig:2} that is obtained using just $J=20$.  Under these practical conditions the relationship between the observed count data, $C_{s}^{\text{o}}(i,k)$ for $i=1,2,3,\ldots,I$, and the solution of the continuum limit PDE, $c_s(x,t)$, is unclear.  We will make progress in relating these two quantities by introducing two different types of measurement error models that account for the fluctuations in the count data in different ways~\cite{murphy2024implementing}.

\subsection{Data} \label{sec:data}
For this simulation study we use synthetic data generated by stochastic model described in Section~\ref{sec:sm} to generate count data that has similar properties to count data obtained in a scratch assay. With this data we will explore different options for efficient likelihood-based parameter estimation, parameter identifiability analysis and model prediction.  We will focus on two different cases: (i) Case 1 involves generating data at a single time point involving biased motility in the context of working with a single homogeneous population with $S=1$; and (ii) Case 2 involves generating data at a single time point involving biased motility with a population composed of two subpopulations with $S=2$,  All data are denoted using the vector $y^{\text{o}}$. For Case 1, $y^{\text{o}}$ is a vector of length $I$; for Case 2, $y^{\text{o}}$ is a vector of length $2I$.  

\begin{figure}[htp!]
\centering
\includegraphics[width=1\textwidth]{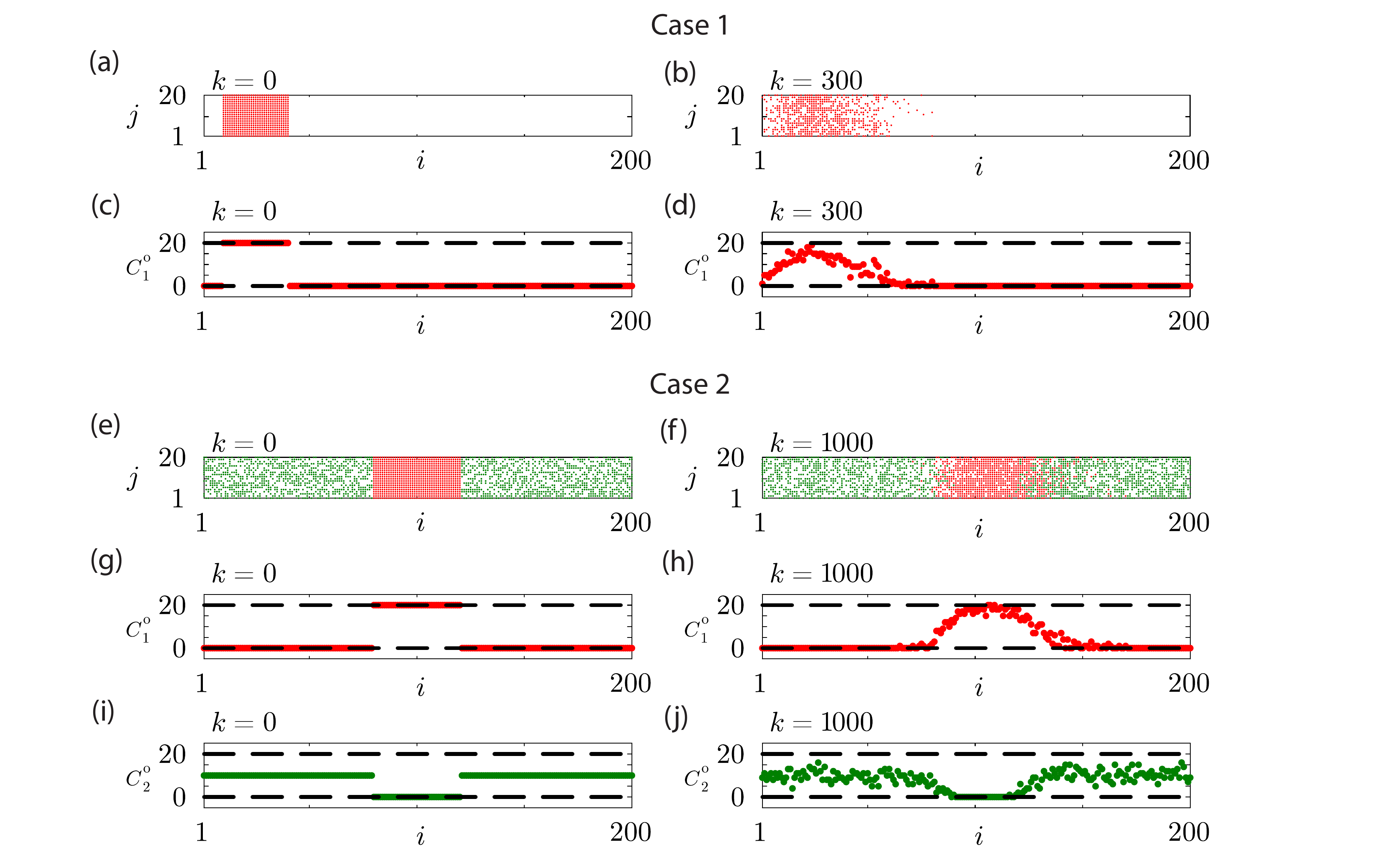}
\caption{Snapshots of agent distribution and associated count data from stochastic simulations. Parameters in the stochastic model are $(P_1, \rho_1)^\top = (1, 0.1)^\top$ for Case 1 and $(P_1, P_2,\rho_1,\rho_2)^\top = (0.8, 1, 0.2, 0)^\top$ for Case 2. (a)-(b) and (c)-(d) show snapshots and count data for Case 1. (e)-(f) shows snapshots for Case 2, whereas (g)-(j) shows count data for Case 2.}
\label{fig:3}
\end{figure}

Case 1 involves a biased single population, $S=1$, on a $200 \times 20$ lattice where the motion of agents is biased, $(P_1,\rho_1)^\top = (1,0.1)^\top$ which corresponds to $(D_1,v_1)^\top = (0.25,0.05)^\top$ in the continuum model under idealised conditions where $J$ is sufficiently large.  Since the motion is biased in the positive $x$-direction, the initial placement of agents involves fully occupying sites with $10 \leq i \leq 40$, as shown in Figure~\ref{fig:3}(a).  The placement of agents towards the left boundary of the domain allows us to observe the biased motion in the positive $x$-direction, and data collected after $k=300$ time steps gives rise to the snapshot shown in Figure~\ref{fig:3}(b). Count data in Figure \ref{fig:3}(c)--(d) are given at the beginning and end of the simulation, respectively. 

Case 2 involves a biased population composed of two subpopulations, $S=2$, on a $200 \times 20$ lattice with $(P_1, P_2, \rho_1, \rho_2)^\top = (0.8, 1, 0.2, 0.0)^\top$, corresponding to $(D_1,D_2,v_1,v_2)^\top = (0.2, 0.25, 0.08, 0.0)^\top$ in the continuum model under idealised conditions where $J$ is sufficiently large. The initial placement of agents involves placing a group of agent from subpopulation 1 in the central region of the lattice so that all sites with $80 \leq i \leq 120$ are completely occupied by agents from subpopulation 1.  All remaining lattice sites are randomly occupied by agents from subpopulation 2 with probability 0.5, as shown in Figure~\ref{fig:3}(e).    This means that agents from subpopulation 1 will attempt to spread out from their original location with a small bias in their motion in the positive $x$-direction.  Since the motion of agents from subpopulation 1 are hindered by the presence of surrounding agents from subpopulation 2 we collect data after a larger number of time steps, $k=1000$, which gives rise to the snapshot in Figure~\ref{fig:3}(f). Count data for subpopulation 1 are given in Figure \ref{fig:3}(g)--(h) whereas count data for subpopulation 2 are given in Figure \ref{fig:3}(i)--(j). 

\newpage
\subsection{Likelihood function}\label{sec:lf}
Our count data, summarised in Figure \ref{fig:3}, are deliberately generated with $J=20$ so that the count data involve clear fluctuations and so the mathematical relationship between the discrete count data and the solution of the corresponding continuum limit model, Equation~\eqref{eq:CtmDiscrete}, is not guaranteed to hold.  Therefore, we introduce two \textit{measurement error models}~\cite{murphy2024implementing} that allow us to relate the solution of the continuum model to the noisy count data in a probabilistic sense.  In the first instance we take a standard approach and work with an additive Gaussian measurement error model because this is the most standard approach in the biological physics and mathematical biology literature.  Secondly, we introduce a multinomial measurement error model which, as we will show, is a natural choice for working with count data.

The standard approach for relating data to solutions of differential equations is to assume that the data $y^{\text{o}}$ is normally distributed about the solution of the differential equation of interest. Throughout this work we will refer to this approach as working with an \textit{additive Gaussian measurement error model}. For example, if we consider Case 1 where we have $y^{\text{o}} = C_{1}^{\text{o}}(i,k)$ for $i=1,2,3,\ldots,I$ and the solution of the continuum limit mathematical model is $c_1(x_i,t)$ for $i=1,2,3,\ldots,I$, we make the standard assumption that $C_{1}^{\text{o}}(i,k) = \mathcal{N}(c_1(x_i,t),\sigma_1^2)$ for $i=1,2,3,\ldots,I$.   Since we deal with $S$ counts across $I$ columns of the lattice, invoking a standard independence assumption gives us a log-likelihood function that can be written as 
\begin{equation}
\label{eq:log_gaussian}
\ell (\theta \, | \, y^{\text{o}}) =  \sum_{s=1}^{S} \sum_{i = 1}^{I} \log \left[ \phi\left(C_{s}^{\text{o}}(i,k)/J \mid  c(x_i,t), \sigma_s^2 \right)\right],
\end{equation}
where $\phi(x \mid \mu, \sigma^2)$ is the Gaussian probability density function with mean $\mu$ and variance $\sigma^2$, and $\theta$ refers to the parameters in the continuum limit PDE model. 

Working with an additive Gaussian measurement error model means that we have introduced additional parameters in the noise model, $\sigma_s^2$ for $s=1,2,3,\ldots,S$.  Throughout this work we will present results generated using the Gaussian additive noise model in terms of the standard deviation $\sigma_s$ for $s=1,2,3,\ldots,S$.  As we will demonstrate, data will be used to estimate the parameters in the continuum limit model description of the stochastic model as well as simultaneously estimating parameters in the measurement error model.  For example, in Case 1 where we have $S=1$ we have two parameters in the continuum limit model $D_1$ and $v_1$, and one parameter in the measurement error model $\sigma_1$.  When working with the additive Gaussian measurement error model we will include both sets of parameters in the vector $\theta$ which means that with $S=1$ we have $\theta = (D_1,v_1,\sigma_1)^\top$ whereas with $S=2$ we have $\theta = (D_1,D_2,v_1,v_2,\sigma_1,\sigma_2)^\top$.

Instead of simply assuming that the data is normally distributed about the solution of the appropriate continuum limit model, we can take advantage of the structure of our count data and derive a more mechanistically-motivated measurement error model that we will refer to as a \textit{multinomial measurement error model}.  For example, when dealing with just a single population, $S=1$, our count data simply consists of counts of agents within each column of the lattice, which implicitly defines a count of the vacant lattice sites in each column, meaning that our data takes the form  $C_1^\text{o}(i,k)$ and $E^\text{o}(i,k) = J - C_1^\text{o}(i,k)$ for $i=1,2,3,\dots,I$.   If we interpret $C_1^\text{o}(i,k)$ at any given spatial location as corresponding to a finite number of independent samples from some underlying stochastic process, these samples can be thought of as an approximation of a continuous measurement of agent occupancy within that column.  These data can be considered as samples from a distribution where the expected occupancy fraction is given by the solution of the continuum limit model $c_s(x_i,t)$.  Together, these measurements imply a binomial likelihood,
\begin{equation}
\label{eq:binomial}
\mathcal{L} (\theta \mid y^\text{o}(i)) \propto c_1(x_i,t)^{C_1^\text{o}(i,k)}\left(1-c_1(x_i,t)\right)^{E^\text{o}(i,k)}.
\end{equation}
Assuming all column counts, $i=1,2,3,\ldots, I$, are conditionally independent, given the continuum-limit solution, and taking the logarithm of the likelihood function gives us a log-likelihood function based on the binomial measurement error model for a single population with $S=1$ that can be written as
\begin{equation}
\label{eq:log_binomial}
\ell  (\theta \mid y^\text{o}) \propto \sum_{i=1}^I \log \left[   c_1(x_i,t)^{C_1^\text{o}(i,k)}\left(1-c_1(x_i,t)\right)^{E^\textrm{o}(i,k)}\right].  
\end{equation}

Similar arguments lead to a multinomial-based log-likelihood function for $S >1$.  Under these conditions our count data  consists of counts of agents from each subpopulation within each column,  $C_s^\text{o}(i,k)$ for $i=1,2,3,\ldots,I$ and $s=1,2,3,\ldots,S$.  The geometry of the lattice means that counts of vacant sites is given by $\displaystyle{E^\text{o}(i,k) = J - \sum_{s=1}^S C_s^\text{o}(i,k)}$ for $i=1,2,3,\dots,I$.   Again, we can interpret $C_s^\text{o}(i,k)$ at any given spatial location as being generated by a finite number of independent samples from some underlying stochastic process, these samples can be thought of as an approximation of expected, noise-free measure of agent occupancy from the $s$th subpopulation within that column.  These data can be considered as samples from a distribution where the expected occupancy fraction is given by the solution of the continuum limit model, $c_s(x_i,t)$.  Following the same arguments as above for $S=1$, these measurements imply multinomial log-likelihood function which can be written as  
\begin{equation}
\label{eq:log_multinomial}
\begin{aligned}
&\ell (\theta \, | y^{\text{o}})\propto &\sum_{i = 1}^{I} \log\left[\prod_{s=1}^{S}c_{s}(x_{i},t)^{C_{s}^{\text{o}}(i,k)}\left(1 - \sum_{s=1}^{S}c_{s}(x_{i},t)\right)^{E^\textrm{o}(i,k)} \right].  
\end{aligned}
\end{equation}
This multinomial log-likelihood function relaxes to the binomial log-likelihood function, derived previously, when $S=1$, and here we have written the data using the compact vector notation $y^{\text{o}}$.  In practice, when we evaluate the multinomial log-likelihood function we simply set the proportionality constant to unity.  Unlike the additive Gaussian measurement error model, working with the multinomial measurement error model does not involve introducing any new parameters.  When we work with $S=1$ we have $\theta = (D_1,v_1)^\top$, whereas $S=2$ involves $\theta = (D_1,D_2,v_1,v_2)^\top$.

\subsection{Likelihood-based estimation and identifiability}
Given a set of count data $y^{\text{o}}$ together with an appropriate process model and measurement error model with all unknown parameters summarised in the vector $\theta$, we have access to a log-likelihood function, $\ell(\theta \, | y^{\text{o}})$.  Consequently, the choice of parameters that gives the best match to the data is given by $\hat{\theta}$ which maximises the log-likelihood, giving rise to the maximum likelihood estimate (MLE),
\begin{equation}
\label{eq:MLE}
\hat{\theta} = \underset{\theta}{\arg \sup} \left[\ell (\theta \, \vert \, y^{\text{o}})\right].
\end{equation}
Throughout this work we always use numerical optimization calculate $\hat{\theta}$.  In particular all numerical optimization calculations are performed using the Nelder-Mead algorithm with simple bound constraints~\cite{NELDERMEAD1965} implemented within the NLopt routine~\cite{NLopt2007} in Julia. In general we find that our numerical optimization calculations are insensitive to the initial estimate and that the algorithm converges using default stopping criteria.

Given our estimate of $\hat{\theta}$, we use then the profile likelihood to quantify the precision of our parameter estimates by examining the curvature of the log-likelihood function~\cite{pawitan2001,raue2009,raue2013,villaverde2019,villaverde2022}.  To compare our results against asymptotic thresholds we work with a normalised log-likelihood function 
\begin{equation}
\label{eq:log-likelihood}
\begin{aligned}
\bar{\ell} (\theta \mid y^{\text{o}}) &= \ell (\theta \mid y^{\text{o}})- \ell (\hat{\theta} \mid y^{\text{o}}),
\end{aligned}
\end{equation}
so that $\bar{\ell} (\hat{\theta}  \mid y^{\text{o}})=0$.  To proceed we partition the full parameter $\theta$ into \textit{interest} parameters $\psi$, and \textit{nuisance} parameters $\omega$, so that $\theta = (\psi, \omega)$.  In this study we restrict our attention to univariate profile likelihood functions which means that our interest parameter is always a single parameter.  For a set of data $y^{\text{o}}$, the profile log-likelihood for the \textit{interest} parameter $\psi$ given the partition $(\psi, \omega)$ is
\begin{equation}
\label{eq:prof-likelihood}
\begin{aligned}
    \bar{\ell}_p (\psi \mid  y^{\text{o}}) =  \sup_{\omega \mid \psi} \left[\bar{\ell}(\psi, \omega \mid y^{\text{o}}) \right],
\end{aligned}
\end{equation}
which implicitly defines a function $\omega^*(\psi)$ of optimal values of $\omega$ for each value of $\psi$. As for calculating the MLE, all profile likelihood functions in this work are calculated using the Nelder-Mead numerical optimization algorithm with the same bound constraints used to calculate $\hat{\theta}$.  As a concrete example, if we work with the continuum limit model for a single population using the additive Gaussian measurement error model we have $\theta = (D_1,v_1,\sigma_1)^\top$.  In this scenario we can compute three univariate profile likelihood functions by choosing the interest parameter to be: (i) the diffusivity, $\psi = D_1$ and $\omega = (v_1,\sigma_1)^\top$; (i) the drift velocity, $\psi = v_1$ and $\omega = (D_1,\sigma_1)^\top$; and (iii) the standard deviation in the measurement error model, $\psi = \sigma_1$ and $\omega = (D_1,v_1)^\top$.  For all univariate profile likelihood calculations we work with a uniformly-discretized interval that contains the MLE.  For example, if our interest parameter is $D_1$ we identify an interval,  $D_1^- <\hat{D}_1 <  D_1^+$ and evaluate $\bar{\ell}_p$ across a uniform discretization  of the interval to give a simple univariate function that we call the profile likelihood.  The degree of curvature of the profile likelihood function provides information about the practical identifiability of the interest parameter.  For example, if the profile log-likelihood function is flat then the interest parameter is non-identifiable.  In contrast, when the profile log-likelihood function is curved the degree of curvature indicates inferential precision and we may determine likelihood-based confidence intervals where $\bar{\ell}_p < -\Delta_{q,n}/2$ where $\Delta_{q,n}$ denotes the $q$th quantile of the $\chi^2$ distribution with $n$ degrees of freedom, which we take to be the relevant number of unknown parameters~\cite{Royston2007}.  For example, identifying the interval where $\bar{\ell}_p < -\Delta_{0.95,1}/2 \approx -1.9207$ allows us to identify the asymptotic 95\% confidence interval~\cite{pawitan2001,murphy2024implementing} for a univariate profile likelihood function with one free parameter.  This procedure gives us a simple way of identifying the width of the interval $[D_1^-, D_1^+]$.  A simple approach to determine suitable choices of $D_1^-$ and $D_1^+$ is to compute $\bar{\ell}_p$ across some initial estimate of the interval and if the values of the profile log-likelihood do not intersect the relevant asymptotic threshold then we simply continue to compute the profile log-likelihood function on a wider interval.  In contrast, if we compute the profile log-likelihood on some interval and find that this interval is very wide compared to the relevant asymptotic threshold we can simply re-compute the profile log-likelihood function on a narrower interval.  Since these computations involve numerical optimisation calculations where we always have a reasonably good estimate to start the iterative calculations (e.g. $\hat{\theta}$) we find that these computations are efficient.

\subsection{Likelihood-based prediction}
The methods outlined so far focus on taking noisy count data and exploring how to find the MLE parameter estimates, and to take an asymptotic log-likelihood threshold to define a parameter confidence set so that we can understand how variability in count data corresponds to variability in parameter estimates.  In this section we will now explore how variability in parameter estimates maps to variability in model predictions.   In particular, our focus will be to compare likelihood-based predictions using the traditional additive noise model with the multinomial noise model and explore differences in these approaches.  

Given a set of count data, $y^{\textrm{o}}$, and an associated normalised log-likelihood function, $\bar{\ell}(\theta \mid y^{\textrm{o}})$, we proceed by identifying the asymptotic 95\% log-likelihood threshold $-\Delta_{0.95,n}/2$, where $n$ is the number of free parameters.  Using rejection sampling we obtain $M$ parameter sets that lie within the 95\% log-likelihood threshold so that we have $\theta_{m}$ that satisfy $\bar{\ell} (\theta_{m} \, | \, y^{\text{o}}) \geq -\Delta_{0.95,n}/2$ for $m=1,2,3,\ldots,M$. For each $\theta_m$ within the confidence set we solve the relevant continuum model to give $M$ solutions that correspond to the \textit{mean trajectory} prediction as defined by the relevant measurement error model~\cite{murphy2024implementing}.  Since all measurement error models considered in this work involve a parametric distribution, we can also construct various \textit{data realization} predictions~\cite{murphy2024implementing} by considering various measures of the width of that parameteric distribution about the mean.  For simplicity we will consider the 5\% and 95\% quantiles of the relevant distributions as a simple measure of distribution width. To make a prediction using the single population model we denote the mean trajectory as $c_1^{(m)}(x,t)$ to denote the solution of Equation~\eqref{eq:cont_1p} using the $m$th parameter sample from within the parameter confidence set.  To plot the solutions we discretize the spatial location so that we have $c_1^{(m)}(x_r,t)$, where $x_r = (r-1)/10$ for $r=1,2,3,\ldots,2001$.  For each of the mean trajectories we now have an interval $c^{(m)}_{1,0.05}(x_r,t)< c_1^{(m)}(x_r,t) < c^{(m)}_{1,0.95}(x_r,t)$ for $r=1,2,3,\ldots,2001$, where the lower bound corresponds to the 5\% quantile of the measurement error noise model and the upper bound corresponds to the 95\% quantile of the measurement error noise model.  For a fixed parameter $\theta_m$, the 5\% and 95\% quantiles of the additive Gaussian measurement error model are constants determined by $\sigma_1$.  For the multinomial measurement error model the 5\% and 95\% quantiles are no longer constants but vary with the mean of the distribution.  For $S=1$ the multinomial distribution relaxes to the binomial distribution and hence the bounds at location $x_r$ correspond to the 5\% and 95\% quantiles of the binomial distribution with $J$ trials with probability of success $c_1(x_r,t)$.  For the multinomial measurement error model with $S > 1$ the bounds for subpopulation $s$ at location $x_r$ correspond to the 5\% and 95\% quantiles of the $s$th marginal distribution, again with $J$ trials with probability of success $c_s(x_r,t)$. After calculating these $M$ intervals around each mean trajectory we then take the union to define $\left[\text{min}\left(c^{(m)}_{1,0.05}(x_r,t)\right), \text{max}\left(c^{(m)}_{1,0.95}(x_r,t)\right)\right]$.  Here the minimum value of $c^{(m)}_{1,0.05}(x_r,t)$ and the maximum value of $c^{(m)}_{1,0.95}(x_r,t)$ are computed for each fixed value of $x_r$ across the set of $M$ different parameter values, $m=1,2,3,\ldots,M$.  This simple computational procedure gives us a complete picture of the prediction interval that can be interpreted as a form of tolerance interval~\cite{vardeman1992other} accounting for both parameter uncertainty and data realization uncertainty~\cite{simpson2024}.

\section{Results and Discussion} \label{sec:Results} 

\subsubsection*{Case 1} 
Working with the additive Gaussian measurement error model the MLE for the count data in Case 1 is $\hat{\theta} = (\hat{D}_{1}, \hat{v}_{1}, \hat{\sigma}_{1})^\top = (0.2613, 0.0480, 0.0598)^\top$.  The estimates of the diffusivity and drift velocity differ slightly from the idealised result of $D_1 = 0.25$ and $v_1 = 0.05$ for perfect noise-free data with sufficiently large $J$.  We attribute these differences to the role of fluctuations in the count data that are obtained with just $J=20$.  When working with the multinomial measurement error model the MLE for the same count data is $\hat{\theta} = (\hat{D}_{1}, \hat{v}_{1})^\top = (0.2509, 0.0454)^\top$.  Overall, comparing the MLE estimates of the diffusivity and drift velocity indicate that the different measurement error noise models lead to small differences in the MLE for $D_1$ and $v_1$.  Here, the only practical difference is that working with the additive Gaussian measurement error model involves additional computational effort required to estimate $\hat{\sigma}_1$.   

The broad similarity between our results for the two measurement error models can be partly explained by noting that the multinomial error model simplifies to a binomial error model when $S=1$, and that the binomial distribution can be approximated by a Gaussian distribution under well-known conditions.  The central limit theorem indicates that $C/J \to \mathcal{N}(c,c(1-c)/J)$ when both $Jc$ and $J(1-c)$ are sufficiently large.  In Case 1 we have $N_1 = 620$ on a lattice of size $200\times20$ which means the average occupancy across all columns of the lattice is $\bar{c}_1 = 620/(200\times20) = 0.155$.  On average this is equivalent to having 2 or 3 agents per column across the lattice.  We will make progress with this estimate of $\bar{c}_1$ despite the fact there is a very high variability in the number of agents per column in Figure \ref{fig:3}(d) where we see that many columns contain zero agents, whereas there is one column containing 19 agents.  Under this clearly questionable approximation, the normal approximation of the binomial distribution implies that $\sigma_1 = \sqrt{\bar{c}_1(1 - \bar{c}_1)/J}$, or $\sigma_1 = 0.08$ for our data.  Despite these assumptions, this estimate of $\sigma_1$ is not dramatically different from our estimate of $\hat{\sigma}_{1} = 0.0598$ obtained by using the additive Gaussian measurement error model directly.  This is consistent with our observations that our parameter estimates are not very sensitive to the choice of measurement error model.

\begin{figure}[htp]
\centering
\includegraphics[width=1\textwidth]{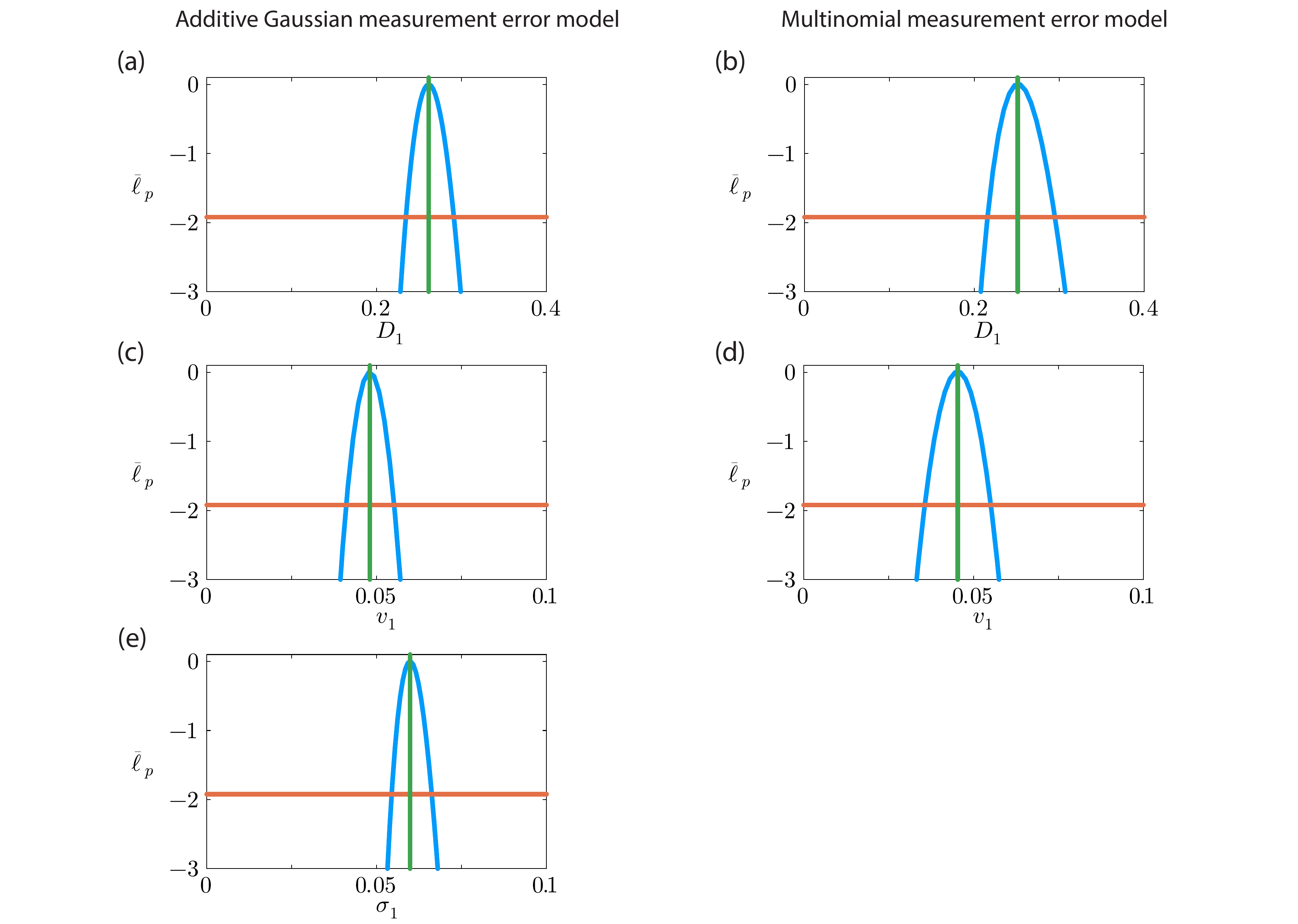}
\caption{Parameter estimation and parameter identifiability for Case 1.  Results in the left column correspond to the additive Gaussian measurement error model, whereas results in the right column correspond to the multinomial measurement error model.  Each subfigure shows a univariate profile likelihood function profile superimposed with a vertical green line at the MLE, and a horizontal orange line at the 95\% asymptotic threshold.  For the additive Gaussian measurement error model the MLE and 95\% confidence intervals are: $\hat{D}_{1} = 0.2613\, \left[ 0.2345,\, 0.2913\right]$; $\hat{v}_{1} = 0.0480\, \left[ 0.0411,\, 0.0553\right]$ and $\hat{\sigma}_{1} = 0.0598\, \left[0.0544,\, 0.0662\right]$. For the multinomial measurement error model we obtain $\hat{D}_{1} = 0.2509\, \left[ 0.2154,\, 0.2950\right]$ and $\hat{v}_{1} = 0.0454\, \left[0.0355,\, 0.0550\right]$.}
\label{fig:4}
\end{figure}

We now consider the identifiability of our parameter estimates by constructing various univariate profile likelihood functions for each parameter, as shown in Figure \ref{fig:4}.   Regardless of which measurement error model is used, overall we see that all univariate profiles are well formed about a single distinct peak at the MLE indicating that all parameters in both loglikelihood functions are practically identifiable.  For all parameters we are able to identify an interval where $\bar{\ell} \ge  - \Delta_{0.95,1}/2 \approx -1.9207$ to define a 95\% asymptotic confidence interval~\cite{Royston2007}.  For example, with the additive Gaussian measurement error model we have $\hat{D}_1 = 0.2613$ and the 95\% confidence interval is $0.2345 \le D_1 \le 0.2913$, indicating that our estimate is reasonably precise.  Since all profile likelihood functions are fairly narrow we conclude that all parameters are reasonably well identified by this data, and again the main difference between working with the additive Gaussian measurement error model and the multinomial measurement error model is the additional computational effort required to compute the profile likelihood for $\sigma_1$ when working with the additive Gaussian measurement error model.

\begin{figure}[htp]
\centering
\includegraphics[width=1\textwidth]{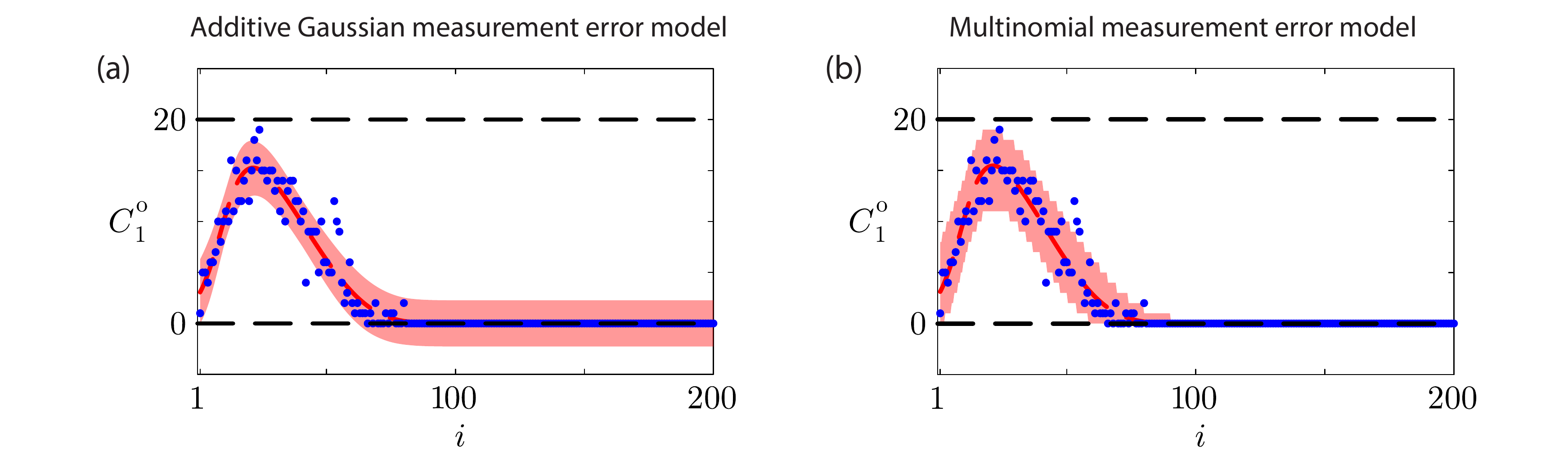}
\caption{Likelihood-based model predictions for data realizations in Case 1. (a) Prediction interval for the additive Gaussian measurement error model; (b) Prediction interval for the multinomial measurement error model.  Each subfigure shows the MLE prediction in a dashed red curve, and a shaded red prediction intervals constructed using $M=500$ parameter values sampled from within the 95\% confidence set for each log-likelihood function.  The 95\% confidence set is associated with  $\bar{\ell} \geq -\Delta_{0.95,3}/2 \approx -3.9074$ for the additive Gaussian measurement error model and $\bar{\ell} \geq -\Delta_{0.95,2}/2 \approx -2.996$ for the multinomial measurement error model.}
\label{fig:5}
\end{figure}

Results in Figure \ref{fig:5} illustrate the prediction intervals for data realizations for Case 1, and here we see a significant difference between the two measurement error models.  Results in Figure \ref{fig:5}(a) for the additive Gaussian measurement error model show that the prediction interval is a smooth interval about the solution evaluated at the MLE, and importantly in regions where the expected counts are zero (i.e. $i \gtrapprox 80$), the lower bound of the prediction interval is negative.  This result is unhelpful because count data are, by definition, non-negative.  This is an important limitation of working with the standard additive Gaussian measurement error model, since the prediction interval does not necessarily obey the physical constraint that $C_1^{\textrm{o}}(i,k) \in [0,J]$.   Repeating this exercise using a different initial arrangement of agents on the lattice, or stopping the simulation at an earlier time can lead to prediction intervals that exceed $J$. In contrast, results in Figure \ref{fig:5}(b) for the multinomial measurement error model show that the prediction interval is not smooth, and this reflects the fact that count data are non-negative integers.  Importantly, the prediction intervals derived using the multinomial measurement error model obey the physical constraint $C_1^{\textrm{o}}(i,k) \in [0,J]$.   Therefore, unlike our results in Figure \ref{fig:4} where the details of the measurement error model made little difference beyond computational efficiency, here in terms of prediction we see that the standard approach can lead to non-physical outcomes whereas the physically-motivated multinomial measurement error model is more computationally efficient to work with, as well as leading to physically meaningful prediction intervals.  For the particular results in Figure \ref{fig:5} we have 94\% of count data lying within the prediction interval for the additive Gaussian measurement error model and 97.5\% of count data falling within the prediction interval for the multinomial measurement error model. 

We will now repeat this exercise for Case 2 to explore the consequences within the context of working with a population composed of multiple interacting subpopulations.

\vfill

\subsubsection*{Case 2}
With the additive Gaussian measurement error model the MLE for the count data in Case 2 is $\hat{\theta} = (\hat{D}_{1}, \hat{D}_2, \hat{v}_{1}, \hat{v}_2, \hat{\sigma}_{1}, \hat{\sigma}_{2})^\top = (0.1774, 0.0971, 0.0717, 0.0024, 0.0639, 0.1094)^\top$.  As for Case 1, these estimates of diffusivities and drift velocities differ from the idealised results. This difference is because we are working with noisy data with $J=20$ as well as the approximate nature of the mean-field PDE model.  When working with the multinomial measurement error model the MLE is $\hat{\theta} = (\hat{D}_1, \hat{D}_2, \hat{v}_1, \hat{v}_2)^\top = (0.1664, 0.0850, 0.0678, 0.0023)^\top$.  Comparing the MLE estimates indicate that the different measurement error noise models only leads to small differences in the MLE, and main practical difference is that working with the additive Gaussian measurement error model involves additional computational effort required to estimate both $\sigma_1$ and $\sigma_2$.   

\begin{figure}[htp]
\centering
\includegraphics[width=1\textwidth]{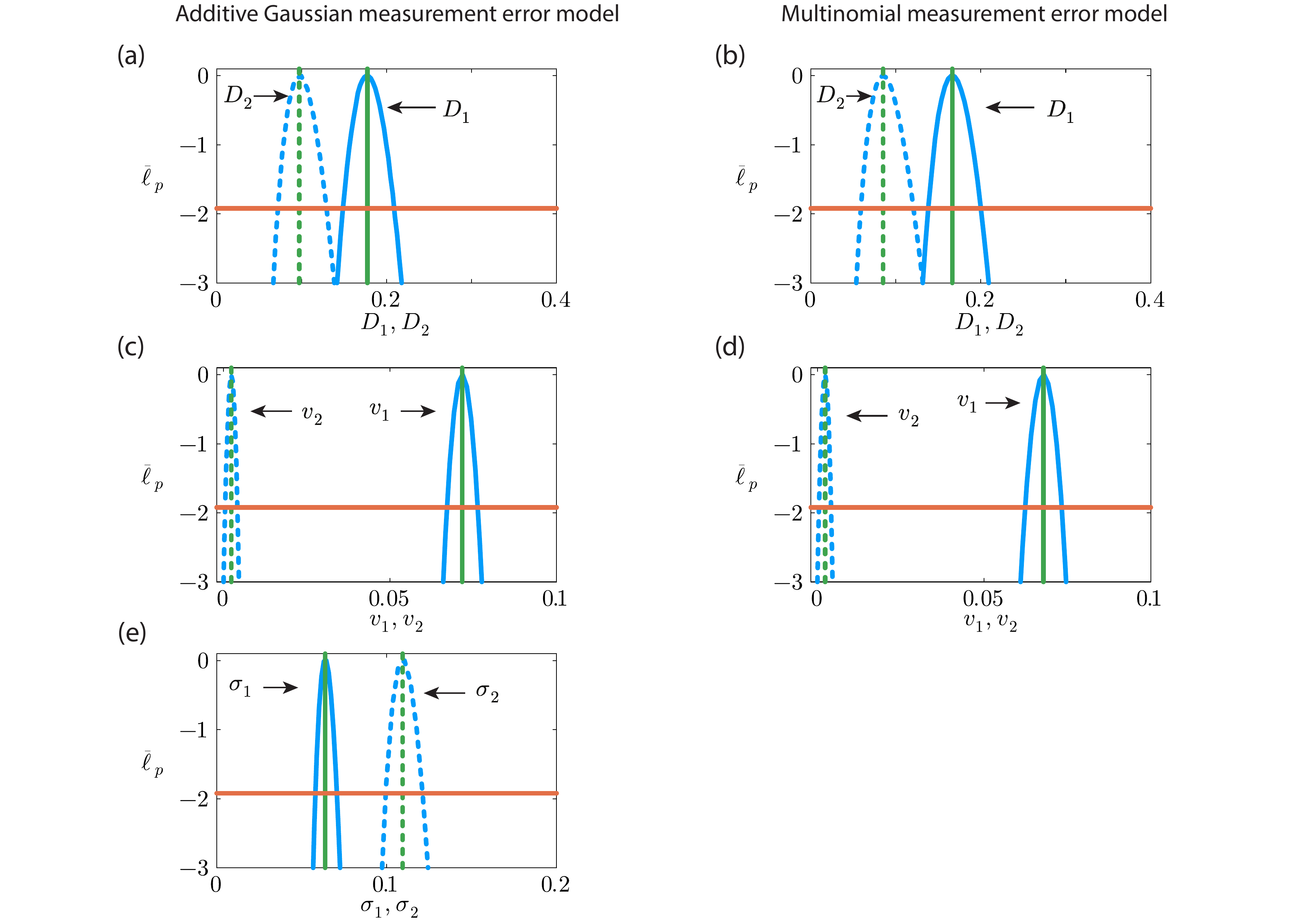}
\caption{Parameter estimation and parameter identifiability for Case 2. Results in the left column correspond to the additive Gaussian measurement error model, whereas results in the right column correspond to the multinomial measurement error model.  Each subfigure shows two univariate profile likelihood functions. Solid lines correspond to subpopulation $s=1$ and dashed lines correspond to subpopulation $s=2$.  Each profile likelihood function is superimposed with a vertical green line at the MLE, and a horizontal orange line at the 95\% asymptotic threshold.   For the additive Gaussian measurement error model the MLE and 95\% confidence intervals are: $\hat{D}_1 = 0.1774\, \left[ 0.1488,\, 0.2089\right]$; $\hat{D}_2 = 0.0971\, \left[ 0.0721,\, 0.1299\right]$; $\hat{v}_1 = 0.0717\, \left[ 0.0671,\, 0.0764\right]$; $\hat{v}_2 = 0.0024\, \left[ 0.0006,\, 0.0043\right]$; $\hat{\sigma}_1 = 0.0639\, \left[0.0580,\, 0.0707\right]$ and $\hat{\sigma}_2 = 0.1094\, \left[0.0995,\, 0.1210\right]$. For the multinomial measurement error model the MLE and 95\% confidence intervals are $\hat{D}_{1} = 0.1664\, \left[ 0.1382,\, 0.2001\right]$; $\hat{D}_{2} = 0.0850\, \left[0.0585,\, 0.1211\right]$; $\hat{v}_{1} = 0.0678\, \left[ 0.0624,\, 0.0734\right]$ and $\hat{v}_{2} = 0.0023\, \left[0.0005,\, 0.0041\right]$.}
\label{fig:6}
\end{figure}

The identifiability of our parameter estimates is explored by constructing various univariate profile likelihood functions shown in Figure \ref{fig:6}.  Again, regardless of which measurement error model is used, all univariate profiles are well formed about a single distinct peak at the MLE indicating that all parameters in both loglikelihood functions are practically identifiable with reasonably narrow 95\% asymptotic confidence intervals. Again the main difference between working with the additive Gaussian measurement error model and the multinomial measurement error model is the additional computational effort required to compute the profile likelihoods for $\sigma_1$ and $\sigma_2$ when working with the additive Gaussian measurement error model.

\begin{figure}[htp]
\centering
\includegraphics[width=1\textwidth]{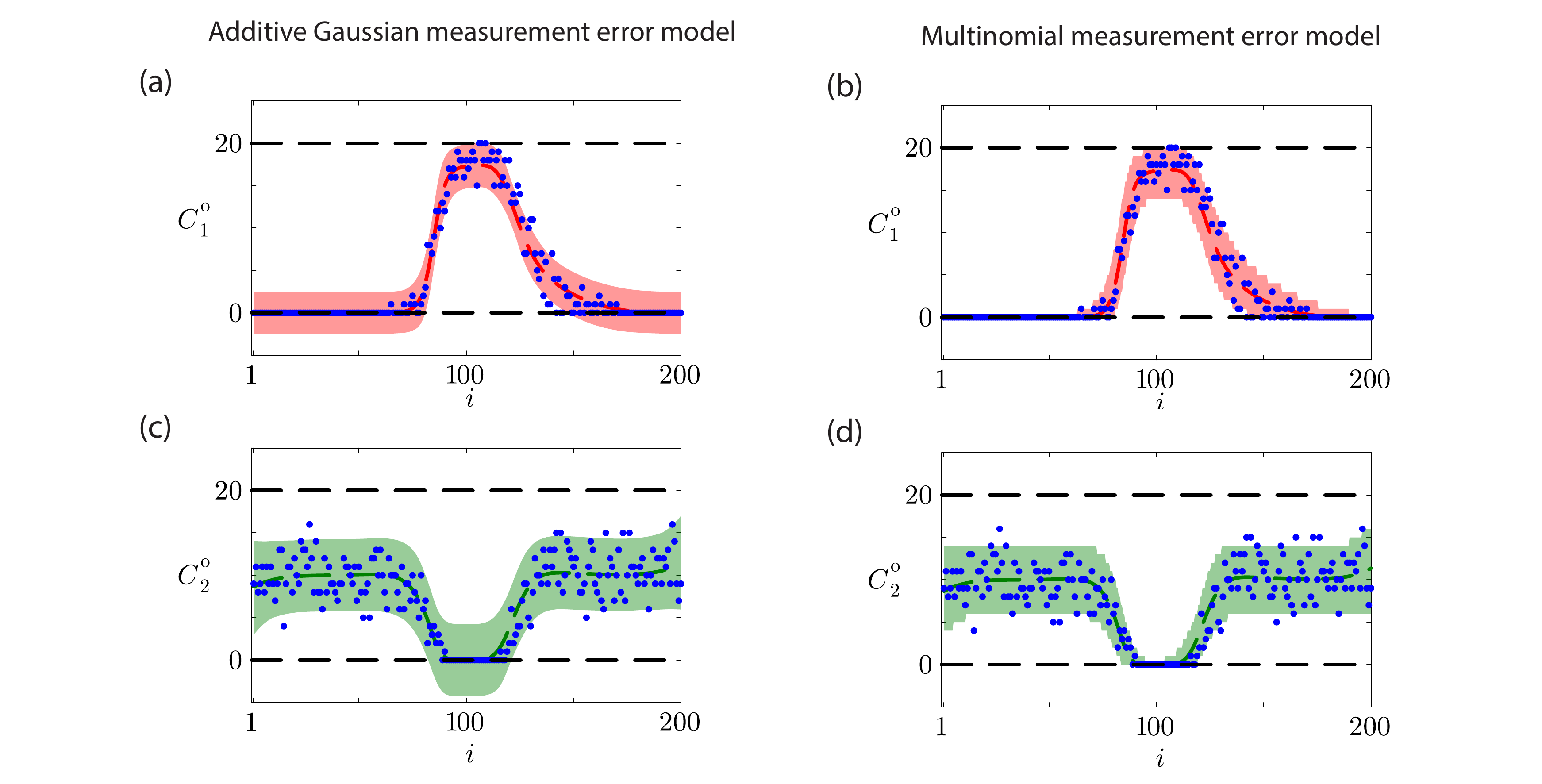}
\caption{Likelihood-based model predictions for data realizations in Case 2. (a) and (c) Prediction intervals for the additive Gaussian measurement error model for $s=1$ and $s=2$, respectively. (b) and (d) Prediction intervals for the multinomial measurement error model for $s=1$ and $s=2$, respectively.  Each subfigure shows the MLE prediction in a dashed red or green curve, superimposed with a shaded red or green prediction interval constructed using $M=500$ parameter values sampled from within the 95\% confidence set for each log-likelihood function.  The 95\% confidence set is associated with $\bar{\ell} \geq -\Delta_{0.95,6}/2 \approx -6.2958$ for the additive Gaussian measurement error model and $\bar{\ell} \geq -\Delta_{0.95,4}/2 \approx -4.7438$ for the multinomial measurement error model.}
\label{fig:7}
\end{figure}

Results in Figure \ref{fig:7} illustrate the prediction intervals for data realizations for Case 2, and again we see a significant difference between the two measurement error models.  Results in Figure \ref{fig:7}(a) and (c) for the Gaussian measurement error model show smooth prediction intervals about the MLE solution, and in regions of low density the lower bound of the prediction interval is negative which is physically impossible.  Alternatively, results in Figure \ref{fig:7}(b) and (d) for the multinomial measurement error model leads to physically realistic, non-smooth predictions that reflect the fact that count data are non-negative integers that obey the physical constraint that $C_1^{\textrm{o}}(i,k) \in [0,J]$ and $C_2^{\textrm{o}}(i,k) \in [0,J]$.   For the particular results in Figure \ref{fig:7} we have 94.5\% of count data lying within the prediction interval for the additive Gaussian measurement error model and 96.75\% of count data falling within the prediction interval for the multinomial measurement error model. 

\vfill
\section{Conclusion and Future Work} \label{sec:Conclusion}
In this work we have presented a likelihood-based framework that can be used for parameter estimation, parameter identifiability analysis and model prediction for lattice-based random walk models. In particular, we focus on lattice-based models of biased migration of $S$ potentially distinct subpopulations of agents that can be relevant for interpreting cell biology experiments that involve multiple interacting subpopulations of cells.  Many recent investigations of parameter estimation for these kinds of stochastic models have focused on various Approximate Bayesian Computation methods based on repeated stochastic simulations, and these approaches are often justified because likelihood models are not always considered.  Here we take a different approach and note that count data from stochastic simulation algorithms can be described using several likelihood models and working with a continuum limit PDE approximation means that we have access to a computationally efficient PDE model together with a simple likelihood model that can facilitate maximum likelihood estimation, profile likelihood-based analysis of parameter identifiability and likelihood-based model prediction.  Our approach can lead to significant computational improvements, for example computing the stochastic simulation data in Figure \ref{fig:3}(a)-(d) for Case 1 is approximately ten times longer than solving the corresponding PDE model parameterised by $\hat{\theta}$ with $\delta = 0.5$, and similar computational improvements hold for Case 2.

Many studies focusing on parameter estimation in the mathematical biology and biophysics literature often work with an additive Gaussian measurement error model to relate noisy observations to the solution of a continuous differential equation.  Here we follow the same approach noting that this approach also leads to additional parameters $\sigma_s$ for $s=1,2,3,\ldots,S$ to be estimated along with the parameters in the process model.   Alternatively, we note that count data obtained from the stochastic simulation model naturally leads to a multinomial measurement error model that can also be implemented and the main contribution of our study is to compare the performance of the likelihood-based parameter estimates, identifiability analysis and model prediction for the two measurement error models.  In general we show that the two approaches can lead to very little differences in terms of parameter estimates and profile likelihood functions, however the outcomes in terms of model predictions is very different.  In particular the standard approach of working with an additive Gaussian measurement error model can lead to non-physical model predictions where the lower bound of the prediction interval is negative and the upper bound of the prediction interval exceeds the maximum packing density of agents per column in the random walk model.  In contrast the multinomial measurement error model leads to physically meaningful prediction intervals that obey physical constraints imposed by the lattice-based modelling framework.

The results presented in this study correspond to working with a stochastic model describing the motion of $S$ subpopulations of agents on a lattice, and the movement of agents in each subpopulation can be biased or unbiased, and this modelling framework can be described by a system of $S$ nonlinear PDEs for the non-dimensional density $c_s(x,t)$ for $s=1,2,3,\ldots,S$.  Our framework can also be applied to a broader set of mechanisms and more complicated PDE models.  For example, incorporating a birth-death process into the stochastic model~\cite{Baker2010} leads to a more complicated stochastic model where $N_s$ can now change over time and the mean-field PDE description involves a source term,
\begin{equation}
\label{eq:cont2}
\frac{\partial c_{s}}{\partial t} = -\frac{\partial \mathcal{J}_{s}}{\partial x} + \mathcal{G}_s, \quad \textrm{for} \quad  s = 1,2,3, \dots, S,
\end{equation}
where $c_{s}(x, t)$ is the dimensionless density of population $s$ at location $x$ and time $t$.  Here $\mathcal{J}_s$ denotes the  flux of subpopulation $s$, and this would remains the same as in \eqref{eq:Js} provided that the discrete motility mechanism remains the same, whereas   $\mathcal{G}_s$ denotes a source/sink term for subpopulation $s$ that models the impact of the birth-death process in the stochastic model, and we note that this source term is related to a generalised logistic growth term.  Our approach for relating noisy count data from the stochastic model to the solution of a mean-field PDE model remains unchanged regardless of whether we consider incorporating a birth-death process.  Clearly it is also possible to generalise the motility mechanism beyond working with the simple biased motility mechanisms explored in this work.  Some interesting generalisations would be to incorporate cell pulling/pushing~\cite{chappelle2019pulling,yates2015incorporating} or cell swapping mechanisms~\cite{noureen2023swapping} into the discrete model.  Since the mean-field descriptions of such generalized mechanisms have been previously derived and validated~\cite{chappelle2019pulling,yates2015incorporating,noureen2023swapping} it is straightforward to incorporate these more detailed mechanisms into the same framework outlined in this study.  It is worth noting some caution is warranted, however, as incorporating additional into the discrete model runs a risk of encountering identifiability issues as the size of the parameter space increases.  Therefore, it is prudent to always construct the univariate profile likelihood functions to ensure that parameter estimates are sufficiently precise before biological mechanisms can be associated with parameter estimates.

A final comment is that we chose to present our Case studies in the typical scenario where data is obtained at one time point only.  This simplification was motivated by the fact that it is fairly common in simple cell biology experiments to image the experiment after one fixed time point.  It is conceptually and computationally straightforward to generalise our approach to work with data collected at $K$ time points simply by summing over these additional time points in the log-likelihood function, $k=1,2,3,\ldots,K$.  For example, the multinomial log-likelihood function for count data collected at $K$ time points generalises to
\begin{equation}
\label{eq:log_multinomial2}
\begin{aligned}
&\ell (\theta \, | y^{\text{o}})\propto &\sum_{i = 1}^{I}\sum_{k = 1}^{K} \log\left[\prod_{s=1}^{S}c_{s}(x_i,t_k)^{C_{s}^{\text{o}}(i,k)}\left(1 - \sum_{s=1}^{S}c_{s}(x_{i},t_k)\right)^{E^\textrm{o}(i,k)} \right],  
\end{aligned}
\end{equation}
where $t_k = k \tau$, and the data vector $y^{\text{o}}$ has length $I(S+K)$.  With this modestly generalised log-likelihood function we can employ the same numerical optimization procedures to calculate $\hat{\theta}$ and the associated profile likelihood functions, and the approach for calculating likelihood-based model predictions remains the same.  A similar generalisation can also be implemented to work with the additive Gaussian measurement error model with data collected at $K$ time points.

An important feature of the modelling presented in this study is that we consider a two-dimensional stochastic model of a scratch assay where the density of agents at the beginning of the simulation is independent of vertical location and remains, on average, independent of vertical location during the simulation.  This simplification is consistent with scratch assay design, and under these conditions we work with one-dimensional noisy count data, $C_s^{\textrm{o}}(i,k)$, that depends upon the horizontal position $x_i=(i-1)\Delta$, and time $t = k \tau$.  For other application we may be interested in genuinely two-dimensional count data and it is straightforward to generate and interrogate this kind of data using the same stochastic simulation model by performing a suite of $Q$ identically-prepared realisations.  If $C_s^q(i,j,k)$ is a binary variable that denotes the occupancy status for subpopulation $s$ at site $(i,j)$ after $k$ time steps in the $q$th identically-prepared realization, then $C_s^\textrm{o}(i,j,k) = \displaystyle{\sum_{q=1}^{Q}C_s^q(i,j,k)}$ is the noisy count data generated by considering $Q$ identically-prepared realizations.  Following the same ideas outlined in this work, we can solve a two-dimensional mean-field PDE model to give $c_s(x,y,t)$ and use either an additive Gaussian measurement error model or a multinomial measurement error model to calculate $\hat{\theta}$, the associated profile likelihood-based confidence intervals, and likelihood-based model predictions.  The key difference is that in the current work the count data is associated with $J$ trials where the expected column occupancy fraction is $c_s(i,k)$ and $J$ is the height of the lattice.  For applications where agent density varies with both vertical and horizontal position the count data is associated with $Q$ trials where the expected site occupancy fraction is $c_s(i,j,k)$ and $Q$ is the number of identically-prepared realizations of the stochastic model.  The same concepts apply to three-dimensional applications. 
\vfill
 
\appendix
\section{Numerical methods}
In this work we consider two different continuum models.  For problems involving one population we generate numerical solutions of Equation~\eqref{eq:cont_1p}, $c_{1}(x,t)$. For problems involving two subpopulations we generate numerical solutions of Equation~\eqref{eq:cont_2p},  $c_{1}(x,t)$ and $c_{2}(x,t)$.  We will now describe how we obtain these numerical solutions.  In brief we use a method-of-lines approach where we spatially discretize the PDE models using standard finite difference approximations for the spatial terms and then solve the resulting system of coupled ordinary differential equations (ODE) in time using the DifferentialEquations.jl package by taking advantage of automatic time stepping and temporal truncation error control~\cite{rackauckas2017}.

To solve Equation~\eqref{eq:cont_1p} on $0 < x < L$ we discretize the spatial terms on a uniform grid with grid spacing $\delta > 0$, such that $c_{1}(x_n,t)=c_{1,n}$ for $ n = 1,2,3, \ldots, N$, where $x_n = (n-1)\delta$. At $x=0$ and $x=L$, corresponding to mesh points $x_1$ and $x_N$, respectively, we impose a zero flux boundary conditions $\mathcal{J}_1 = \mathcal{J}_N= 0$ to give
\begin{align}
\dfrac{\text{d}c_{1,1}}{\text{d}t} &= \dfrac{D_{1}}{\delta^2}\left(c_{1,2}-c_{1,1}\right) - \dfrac{ v_{1}}{\delta} \left[c_{1,2}\left(1-c_{1,2}\right)\right],\label{eq:A1} \\
\dfrac{\text{d} c_{1,n}}{\text{d} t}&= \dfrac{D_{1}}{\delta^2}\left(c_{1,n+1} - 2c_{1,n} + c_{1,n-1}\right) \notag \\
&- \dfrac{v_{1}}{2\delta}\left[c_{1,n+1}(1-c_{1,n+1}) - c_{1,n-1}(1-c_{1,n-1})\right], \quad \text{for }  n=2,3,4,\ldots, N-1, \label{eq:A2}\\
\frac{\text{d}c_{1,N}}{\text{d}t} &= -\dfrac{D_{1}}{\delta^2}\left(c_{1,N}-c_{1,N-1}\right) + \dfrac{v_{1}}{\delta} \left[c_{1,N-1}\left(1-c_{1,N-1}\right)\right].\label{eq:A3}
\end{align}
We solve the system of ODEs given by Equations~\eqref{eq:A1}--\eqref{eq:A3} using Heun's method with adaptive time-stepping~\cite{rackauckas2017}. All results presented in this work correspond to $\delta = 0.5$.  To ensure our results are grid independent we considered a number of test cases and checked that numerical results with $\delta=0.5$ were indistinguishable from results with $\delta = 0.1$.

To solve Equation~\eqref{eq:cont_2p} on $0 < x < L$ we discretize the spatial terms on a uniform grid with grid spacing $\delta > 0$, such that $c_1(x_n,t)=c_{1,n}$ and $c_2(x_n,t)=c_{1,n}$ for $ n = 1,2,3, \ldots, N$. No flux boundary conditions are imposed at $x=0$ and $x=L$, leading to
\begin{align}
\dfrac{\text{d}c_{1,1}}{\text{d}t} &= \dfrac{D_{1}}{\delta^2}\left[\left(1-c_{1,2}-c_{2,2}\right)\left(c_{1,2}-c_{1,1}\right) + c_{1,2}\left(c_{1,2}+c_{2,2}-c_{1,1}-c_{2,1} \right)\right] \notag\\ 
 &-\dfrac{v_{1}}{\delta}\left[c_{1,2}\left(1-c_{1,2}-c_{2,2}\right)\right], \notag\\
\dfrac{\text{d}c_{2,1}}{\text{d}t} &= \dfrac{D_{2}}{\delta^2}\left[\left(1-c_{1,2}-c_{2,2}\right)\left(c_{2,2}-c_{2,1}\right) + c_{2,2}\left(c_{1,2}+c_{2,2}-c_{1,1}-c_{2,1} \right)\right] \notag\\
&- \dfrac{v_{2}}{\delta}\left[ c_{2,2}\left(1-c_{1,2}-c_{2,2}\right)\right] , \label{eq:A4}\\
\dfrac{\text{d}c_{1,n}}{\text{d}t} &= \dfrac{D_{1}}{2\delta^2}\left[\left(2-c_{1,n+1}-c_{2,n+1}-c_{1,n}-c_{2,n}\right)\left(c_{1,n+1}-c_{1,n}\right)\right] \notag\\
&+\dfrac{D_{1}}{2\delta^2}\left[\left(c_{1,n+1}+c_{1,n}\right)\left(c_{1,n+1}+c_{2,n+1}-c_{1,n}-c_{2,n}\right)\right] \notag\\
&-\dfrac{D_{1}}{2\delta^2}\left[\left(2-c_{1,n}-c_{2,n}-c_{1,n-1}-c_{2,n-1}\right)\left(c_{1,n}-c_{1,n-1}\right)\right]\notag\\
&- \dfrac{D_{1}}{2\delta^2}\left[\left(c_{1,n}+c_{1,n-1}\right)\left(c_{1,n}+c_{2,n}-c_{1,n-1}-c_{2,n-1}\right)\right] \notag\\ 
&-\dfrac{v_{1}}{2\delta}\left[c_{1,n+1}\left(1-c_{1,n+1}-c_{2,n+1}\right) - c_{1,n-1}\left(1-c_{1,n-1}-c_{2,n-1}\right)\right] \notag\\
\dfrac{\text{d}c_{2,n}}{\text{d}t} &= \dfrac{D_{2}}{2\delta^2}\left[\left(2-c_{1,n+1}-c_{2,n+1}-c_{1,n}-c_{2,n}\right)\left(c_{2,n+1}-c_{2,n}\right)\right]\notag\\
&+ \dfrac{D_{2}}{2\delta^2}\left[\left(c_{2,n+1}+c_{2,n}\right)\left(c_{1,n+1}+c_{2,n+1}-c_{1,n}-c_{2,n}\right)\right] \notag\\
&-\dfrac{D_{2}}{2\delta^2}\left[\left(2-c_{1,n}-c_{2,n}-c_{1,n-1}-c_{2,n-1}\right)\left(c_{2,n}-c_{2,n-1}\right)\right]\notag\\
&-\dfrac{D_{2}}{2\delta^2}\left[\left(c_{2,n}+c_{2,n-1}\right)\left(c_{1,n}+c_{2,n}-c_{1,n-1}-c_{2,n-1}\right)\right] \notag\\ 
&-\dfrac{v_{2}}{2\delta}\left[c_{2,n+1}\left(1-c_{1,n+1}-c_{2,n+1}\right) - c_{2,n-1}\left(1-c_{1,n-1}-c_{2,n-1}\right)\right], \notag\\
&\text{for } n=2,3,4,\ldots, N-1, \label{eq:A5}\\
\dfrac{\text{d}c_{1,N}}{\text{d}t} &= -\dfrac{D_{1}}{\delta^2}\left[\left(1-c_{1,N-1}-c_{2,N-1}\right)\left(c_{1,N}-c_{1,N-1}\right)\right]\notag\\
&-\dfrac{D_{1}}{\delta^2}\left[c_{1,N-1}\left(c_{1,N}+c_{2,N}-c_{1,N-1}-c_{2,N-1}\right)\right]\notag\\
&+ \dfrac{v_{1}}{\delta}\left[c_{1,N-1}\left(1-c_{1,N-1}-c_{2,N-1}\right)\right], \notag\\ 
\dfrac{\text{d}c_{2,N}}{\text{d}t} &= -\dfrac{D_{2}}{\delta^2}\left[\left(1-c_{1,N-1}-c_{2,N-1}\right)\left(c_{2,N}-c_{2,N-1}\right)\right]\notag\\
&-\dfrac{D_{2}}{\delta^2}\left[c_{2,N-1}\left(c_{1,N}+c_{2,N}-c_{1,N-1}-c_{2,N-1}\right)\right]\notag\\
&+ \dfrac{v_{2}}{\delta}\left[ c_{2,N-1}\left(1-c_{1,N-1}-c_{2,N-1}\right)\right] \label{eq:A6}.
\end{align}

We solve the system of ODEs given by Equations~\eqref{eq:A4}--\eqref{eq:A6} using Heun's method with adaptive time-stepping~\cite{rackauckas2017} with $\delta = 0.5$.  Again, we tested these results to ensure they are grid independent. 

\vfill
\printbibliography

\end{document}